\documentclass[twocolumn,amsmath,amssymb]{revtex4}
\usepackage[colorlinks=true, pdfstartview=FitV, linkcolor=blue, citecolor=red, urlcolor=magenta]{hyperref}
\usepackage{amssymb}
\usepackage{latexsym}
\usepackage{epsfig}
\usepackage{color}
\allowdisplaybreaks
\begin{document}
\title{\textbf{Testing new massive conformal gravity with the light deflection by black hole}}
\author{Muhammad Yasir}
\email{yasirciitsahiwal@gmail.com}\affiliation{Department of Mathematics, Shanghai University and Newtouch Center for Mathematics of Shanghai University,  Shanghai ,200444, P.R.China}
\author{Xia Tiecheng}
\email{xiatc@shu.edu.cn}\affiliation{Department of Mathematics, Shanghai University  and Newtouch Center for Mathematics of Shanghai University,  Shanghai ,200444, P.R.China}
\author{Farzan Mushtaq}
\email{farzanmushtaq9@gmail.com}\affiliation{Department of Mathematics, Shanghai University  and Newtouch Center for Mathematics of Shanghai University,  Shanghai ,200444, P.R.China}
 \author{Kazuharu Bamba}
\email{bamba@sss.fukushima-u.ac.jp}\affiliation{Faculty of Symbiotic Systems Science\\
  Fukushima University, Fukushima 960-1296, Japan}
\begin{abstract}
We study the weak gravitational lensing effect of Non-Bocharova-Bronnikov-Melnikov-Bekenstein (BBMB) black hole in new massive conformal gravity. We analyze the deflection angle of light caused by new massive conformal gravity by using the gauss-bonnet theorem. As a consequence, we obtain the gaussian optical curvature and calculate the deflection angle of the new massive conformal gravity for spherically balanced spacetime with the gauss-bonnet theorem. The resultant deflection angle of light in the weak field limits showing that the bending of light is a global and topological phenomenon. Furthermore, we identify the deflection angle of light in the framework of the plasma medium, we also demonstrate the effect of a plasma medium on the deflection of light by non BBMB. In addition, the behavior of the deflection angle by new massive conformal gravity is explicitly shown in the influence of plasma medium and for the non plasma medium.\\
{\bf Keywords:} Weak gravitational lensing; new massive conformal gravity; deflection angle; Gauss-Bonnet theorem.
\end{abstract}
\maketitle

\section{Introduction}
The notable theory of general relativity predicts that light will be bent if an object with a specific gravitational field is intervened in light way and light spreads in unfilled space along a straight line. Gravitational lensing has used to concentrate galaxies, quasars, super-massive black holes, tracing major problems for early stage gravitational waves, and so forth \cite{1}.
In $1801$, the bending point of light by utilizing newtonian mechanics initially determined by Soldner \cite{2}. Einstein determined the comparative Soldner outcome in $1911$, by applying the Minkowski metric \cite{3,4}. It was shown that the spacetime might theoretically resist evaporation in the presence of $f(R)$ gravity. According to standard models of black holes in the Einstein gravity the black hole horizon radius shrinks in vacuum because it loses energy via the Hawking radiation \cite{4*}. Tsujikawa study the inflation, dark energy, local gravity restrictions, cosmic perturbations and spherically symmetric solutions in weak and strong gravitational backgrounds. Also discuss a variety of $f(R)$ theory applications to cosmology and gravity \cite{4**}. This is the starting period of our advanced gravitational lensing. In 1915, Einstein calculated the novel sun rays light as a result of the influence of the spacetime geometry was double that of the main mechanism \cite{5,6}.\\

The concept of massive conformal gravity (MCG) was proposed as a model for massive gravity whose action is composed of the Weyl term and the conformally linked scalar to the Einstein-Hilbert term (CCSE) \cite{1.1, 1.2}. With respect to conformal transformations, this action is invariant. The non propagation of the linearized Ricci scalar is supposed as a strong condition to obtain a massive gravity theory at the linearized level. The new massive conformal gravity (NMCG) is the consequence of the insertion of $R$ breaking conformal symmetry in this direction. A well-known operation that results in the BBMB black hole with a conformal scalar hair is the exclusion of the Weyl term from the NMCG \cite{1.3, 1.5, 1.6, 1.7}. Additionally, the Schwarzschild black hole does not have a continuous limit in the BBMB solution, demonstrating a conformably connected scalar to the metric feature. However, the $R+$CCSE theory will produce a non-BBMB solution with two charges of ADM mass $M$ and scalar charge $Q_{s}$ when the Weyl component is included. The conformal scalar hair now serves as the principal one. We highlight that the non-BBMB solution is the one that the Weyl term switches to. The non-BBMB black hole is a numerical solution to the NCMG, whereas the BBMB black hole is an analytical solution to the NMCG without the Weyl term. It is obvious that the non-BBMB solution is where the Weyl term changes from the BBMB solution. The conformal scalar hair is primary in the non-BBMB solution because the scalar charge is applied separately. The main goal of this paper is to figure out the deflection angle of photon by non-BBMB black hole using Gauss-Bonnet theorem. Our obtained results in the weak-field approximation is significant because it measures a light ray's deflection in a domain outside of the lensing region. This suggests that gravitational lensing has an overall effect, causing multiple light rays to converge between the source and the observer \cite{1.8, 1.9, 1.10, 1.11, 1.12, 1.13, 1.14}.

Einstein predictions during the sunlight based over shadowing affirmed by Eddington in 1919 \cite{6,7}, by considering the dark fluid coupled with dark matter. It is shown that the future dark period for a power-law equation of state may be singularity depending on the parameter choice. Some of the soft future singularities may be avoided by the coupling with dark matter. However, in a generic instance, such linkage does not eliminate the future singularity. Gravitational lensing impact can be seen in \cite{8,9,9*}.
The phenomenon of black holes has been more fascinating in exposing the important physical aspects since the discovery of cosmic acceleration. In general relativity, there are two main categories of vacuum black hole solutions: like uncharged Schwarzschild black hole and charged black hole for example Reissner-Nordstrom black hole. Over the years, several scholars have closely examined these black holes. Many authors study the gravitational lensing caused by various astrophysical phenomena including wormholes, black holes with naked singularities and a few more related things \cite{kb1,kb2,kb3}.

First case of gravitational lensing in which numerous pictures of a double quasar \cite{10,11} found by Walsh, Weymann and Carswell utilized Zwicky in $1979$. Darwin processed the light diversion point because of a solid gravitational field applying the Schwarzschild metric in 1959 \cite{12}.\\
Virbhadra and Ellis contribute critical work in the avoidance edge and powers for the pictures shaped of elliptic integrals of the principal kind due to Schwarzschild black hole \cite{13,14,15}. The development of large-scale structure and the interaction between matter and gravitational potentials are both affected by weak gravitational lensing. The findings of weak lensing can give information because the development of matter disturbances and gravitational potentials differs \cite{15*}. Planck results may be explained by inflationary theories, the spectral index of the scalar modes of the density perturbations and the tensor to scalar ratio by Bamba \cite{15**}. By assuming the Schwarzschild black hole for the solid gravitational lensing, discover the focal point condition and new strategy to discover twisting point. They also tried to solve the lensing problem analytically for the galactic super-massive black hole \cite{16,17,18,18*}.

While gravitational lensing with Schwarzschild black hole with respect to solid field breaking point and twisting edge was additionally figured closely resembling the weak field restrictions. For example, relativistic images, basic bending formulas and weak field limit were also calculated \cite{32,19}. The solid lensing by a normalized symmetric black hole for un-ending grouping of more prominent pictures are shaped  in ref. \cite{20}. After that he broadened his work for black hole. Rindler and Ishak first examinations about cosmological consistent relativistic twisting edge and they indicated that cosmological steady for a Schwarzschild de-Sitter calculation, does not add to the twisting point of light \cite{21,22,23,24,25,25*}.

The gravitational lensing hypothesis explained several wonders involving the gravitational reconfiguration of light. Gravitational lensing causes several impacts including amplification, altered image, distortion, and time delay  \cite{26,28,30,31,33,34,35,39}. Present days gravitational lensing is a significant astrophysical device for discover of in accessible objects, a dispersion of dim issue and higher scale structure, checking of the general relativity disclosure of the globe and current astronomical microwave \cite{40,41,42,43,44,45,46,46a,47,48,50,51,53,54,55,56,57,58} . In this paper we examine presence of plasma impact on different impacts of gravitational lensing.
Light beams travel through plasma in space. In plasma medium, photons go through in numerous impacts like refraction and dispersing. Change the point of redirection of a light beam which is generally utilized in the estimation of mathematical optics through gravitational lensing. It is self-evident, realized that light beams in an inhomogeneous and straightforward medium expand along bending paths \cite{59,60,61,62,63,64,65,66,67}. This phenomenon, which we previously referred to as refraction is notable in everyday life.

According to this explanation, one should look at other astrophysical topics, such as those on wormholes, dark singularities, and naked singularities. They should also look at more references \cite{69,70,71,72,73,74,75,76,77,79,80,81,82}. In \cite{82*}, thoroughly examined at spherically symmetric and black hole solutions, charged black hole solutions, cylindrical solutions, wormhole solutions, and other astro-physically relevant subjects. This topic is crucial because it reveals the underlying workings of the theory and allows for comparisons with other gravitational modifications. In \cite{82**}, focus on the scenario when there is just one fluid, which in general relativity corresponds to dark energy.
In this way, it is essential to consider the gravitational lensing impact framed by topological abandons in the weak gravitational field and further methodologies was first given by Gibbons-Werner \cite{42} to discover the avoidance point which show the world wide properties. The Gibbons-Werner strategy has been applied by various creators for a physical application for example, Werner to fixed black hole and furthermore Jusufi \cite{43} and others authors to discover the quantum adjusted avoidance edge
for Schwarzschild black hole just as wormholes and the other cosmological steady impact \cite{84,85,kb86}. Sakalli and Ovgun \cite{44} as of late acquire application for the Rindler modified \cite{45}.
In this article, we use the gauss bonnet theorem (GBT) to identify the topological surrenders space season light divergence point.
Additionally, the GBT which connects the topological surfaces. To begin with,
one can use the subset-arranged surface area as $(D, x, g)$ to estimate the curvature $K$ by using the Riemannian metric
$g$ and the Euler characteristic $x$. The gauss bonnet theorem is defined as
\begin{equation}
\int\int_{\mathcal{G}_{R}}\mathcal{K}dS+\oint_{\partial
\mathcal{G}_{R}}kdt+\alpha_{t}=2\pi\mathcal{X}(\mathcal{G}_{R}).
\end{equation}
Where $k$ shows the geodesic curvature for $\partial D : {s} \rightarrow D$ and  $\theta_{j}$
represents the outside angle along the $j^{th}$ vertex.
we obtain the gaussian curvature as follow \cite{28},
\begin{equation}
\alpha=-\int\int_{D\infty}\mathcal{K}dS,\nonumber\\
\end{equation}
here, $\mathcal{K}$ and $dS$ denotes the optical curvature and the surface element respectively.
As of late, Gallo and crisnejo \cite{33} discuss about the diversion of light in the plasma medium. After that the approach for GBT for various black holes is continually developed via weak gravitational lensing.  In \cite{46a}, Myunga and Zou studied the non-BBMB black hole solution by solving the equations numerically. They also discussed the BBMB black hole solution with secondary hair is not the same as this numerical solution with primary hair, which is important.
We explored at how the proposed deflection angle demonstrates that gravitational lensing might be a useful technique to investigate the characteristics of black hole singularities. Future astrophysical observations may provide information about how hair affects deflection angle. Any hair finding would be a significant signal beyond of general relativity \cite{ 1.13, 1.14}.
The organization of this paper as follows. In Sec. I, we explain the background of new massive conformal gravity of black hole and also discuss the method for the hawking evaporation. In Sec. II and IV, we calculate the deflection angle of new massive conformal gravity of black hole by utilizing the gauss bonnet theorem in non-plasma and plasma medium respectively. In Sec. III and V, we notice that the graphical impact on deflection angle with respect to new massive conformal gravity. Finally, conclusions and further outlooks are presented in Sec. VI.
\section{Computation Of weak lensing new Massive conformal Gravity}
The new massive conformal gravity metric in spherically coordinate is provided as \cite{46a, 47}
\begin{equation}
ds^{2}=-A(r)dt^{2}+\frac{dr^{2}}{B(r)}+{r}^{2}(d\theta^{2}+\sin^{2}\theta d\phi^{2}).
\end{equation}
$A(r)$ and $B(r)$ represents here
\begin{equation}
A(r)=1-\frac{2M}{r}+\frac{Q_{s}^2}{r^2}+\frac{Q_{s}^{2}(-M^2+Q_{s}^2+\frac{6}{m_{2}^2})}{3r^4}+..\nonumber\\
\end{equation}
\begin{equation}
B(r)=1-\frac{2M}{r}+\frac{Q_{s}^2}{r^2}+\frac{2\underline{}Q_{s}^{2}(-M^2+Q_{s}^2+\frac{6}{m_{2}^2})}{3r^4}+..\nonumber\\
\end{equation}
In the equatorial plane, optical spacetime is written as ($\theta=\frac{\pi}{2}$) to find null geodesics ($ds^2=0$)
\begin{equation}
dt^{2}=\frac{dr^{2}}{A(r)B(r)}+\frac{{r}^{2}{d\phi^2}}{A(r)}.
\end{equation}
By using Eq $(1.2)$ we attain non-zero christopher symbols which is defined as
\begin{eqnarray}\nonumber
\Gamma^{0}_{00}&=&-\frac{B(r)^\prime}{2B(r)}-\frac{A(r)^\prime}{2A(r)},\\  \nonumber
 \Gamma^{0}_{11}&=&\frac{A(r)^{\prime}B(r)r^{2}}{2A(r)}-rB(r),\\
 \Gamma^{1}_{10}&=&\Gamma^{1}_{01}=\frac{1}{r}-\frac{A(r)^{\prime}}{2A(r)}.
\end{eqnarray}
Now, we utilize the appropriate optical metric for which the non-zero christopher symbol was used to find the Ricci Scalar:
\begin{eqnarray}
\mathcal{R}&=&\frac{3A^{\prime}(r)B^{\prime}(r)}{4}-\frac{3A(r)B^{\prime}(r)}{2r}-\frac{A^{\prime}(r)B(r)}{4}
\nonumber\\&&+\frac{2A^{\prime}(r)B(r)}{r}+\frac{B^{\prime}(r)}{2A(r)}.
\end{eqnarray}
The gaussian curvature which is expressed as follows:
$$
\mathcal{K}=\frac{\mathcal{R}}{2}\nonumber.\\
$$
Gaussian curvature is calculated as follows after certain calculations:
\begin{eqnarray}  \nonumber
\mathcal{K}&=&\frac{3A^{\prime}(r)B^{\prime}(r)}{8}-\frac{3A(r)B^{\prime}(r)}{4r}\\
&-&\frac{A^{\prime}(r)B(r)}{8}
+\frac{A^{\prime}(r)B(r)}{r}+\frac{B^{\prime}(r)}{4A(r)}.
\end{eqnarray}
Where $A({r})$, $B(r)$
\begin{equation}
A(r)=1-\frac{2M}{r}+\frac{Q_{s}^2}{r^2}+\frac{Q_{s}^{2}(-M^2+Q_{s}^2+\frac{6}{m_{2}^2})}{3r^4}+..\nonumber\\
\end{equation}
\begin{equation}
B(r)=1-\frac{2M}{r}+\frac{Q_{s}^2}{r^2}+\frac{2\underline{}Q_{s}^{2}(-M^2+Q_{s}^2+\frac{6}{m_{2}^2})}{3r^4}+..\nonumber\\
\end{equation}
so gaussian curvature is stated as
\begin{eqnarray}
\mathcal{K}&=&\frac{5M^{2}}{{2r}^{4}}-\frac{9MQ^{2}}{2r^{5}}-\frac{36MQ^{2}}{r^{7}m^{2}_{2}}-\frac{M}{r^{3}}
-\frac{Q^{2}}{r^{4}}\nonumber\\
&-&\frac{4M^{2}Q^{2}}{3r^{6}}+\frac{12Q^{2}}{r^{6}m^{2}_{2}}+\frac{M}{4r^{4}}
+\frac{3M^{2}}{2r^{3}}-\frac{11MQ^{2}}{4r^{4}}\nonumber\\
&-&\frac{Q^{2}}{4r^{3}}+\frac{Q^{2}M^{2}}{6r^{5}}-\frac{4Q^{2}}{r^{5}m^{2}_{2}}-\frac{8MQ^{2}}{r^{6}m^{2}_{2}}. \label{AH6}
\end{eqnarray}
\subsection{Deflection angle}
For multiple images, we apply the global theory (Gauss-Bonnet theorem) to associate with the local feature of the space-time that corresponds to Gaussian optical curvature  \cite{1.8, 1.9, 1.10, 1.11, 1.12, 1.13, 1.14}. The aforementioned equation will be used to compute the deflection angle by using a non-singular domain $\mathcal{G}_{R}$ outside of the light beam along with boundaries $\partial\mathcal{G}_{R} = \gamma_{g} \cup C_{R}$, Gaussian curvature $K$, geodesic curvature and exterior jump angles $\alpha_{i}= (\alpha_{0}, \alpha_{\mathcal{G}})$ at vertices. Then, the GBT can be stated as follows
\begin{equation}\nonumber
\int\int_{\mathcal{G}_{R}}\mathcal{K}dS+\oint_{\partial
\mathcal{G}_{R}}kdt+\alpha_{t}=2\pi\mathcal{X}(\mathcal{G}_{R}).
\end{equation}
We can use the condition $r = b/ sin\phi$ at zero order in the weak field regions as the light beam corresponds to a straight line,
\begin{equation}\nonumber
\alpha=-\lim_{R\rightarrow 0}\int_{0} ^{\pi} \int_\frac{b}{\sin\phi} ^{R} \mathcal{K} dS.
\end{equation}
Thus, deflection angle up to leading order term become:
\begin{eqnarray}\nonumber
\alpha&=&\frac{M^{2}Q^{2}\pi}{8b^{4}m^{2}_{2}}-\frac{M\pi}{16b^{2}}+\frac{11MQ^{2}\pi}{16b^{2}}-\frac{2Q^{2}M^{2}}{27b^{2}}\\ \label{A1}
&+&\frac{2Q^{2}\pi}{3b^{3}m^{2}_{2}}+\frac{3MQ^{2}\pi}{4b^{4}m^{2}_{2}}+\frac{Q^{2}\pi}{4b^{2}}+\frac{18MQ^{2}\pi}{5b^{2}m^{2}_{2}}\nonumber\\
&+&\frac{2MQ^{2}}{b^{3}}-\frac{5M^{2}\pi}{8b^{2}}+\mathcal{O}(M^3,Q_{s}^3,b^3).
\end{eqnarray}
\section{ Graphical Influence on Deflection angle new Massive conformal Gravity}
The graphical impact of the deflection angle on new massive conformal gravity is discuss in this section of the article.
We determine how various parameters affects on the outcome
angle. To avoid the complexity  we suppose $Q_{s}=Q$ and $m_{2}=m$
\begin{figure}
\includegraphics[width=16pc]{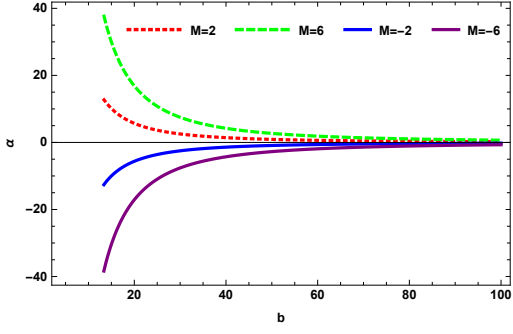}
\caption{\label{label} $\alpha$ versus $b$ with fix value of $m$ and $Q$ for non plasma medium.}
\end{figure}
\begin{figure}
\includegraphics[width=16pc]{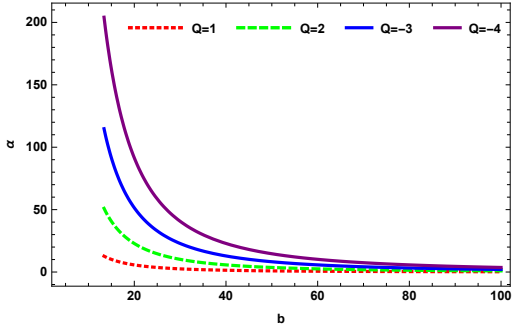}
\caption{\label{label} $\alpha$ versus $b$ with fix value of $m$ and $M$  for non plasma medium.}
\end{figure}
In Fig. 1, we show that the behavior of the deflection angle versus $b$ for fixed value of $m=0.1$, $Q=1$ and by changing $M$. It is to be observed that for $M>0$ the behavior of deflection angle gradually increases. We also analyze that, for $M<0$ the deflection angle decreases negatively and for $M=0$ the deflection angle is linear. In Fig. 2, we show the behavior of deflection angle versus $b$ by fluctuating the charge $Q$ and by taking $M=2$ and $m=0.1$ fixed.
We notice that for the value of $Q>0$ and $Q<0$ the deflection angle gradually expanding.
\section{Weak Lensing By new Massive conformal Gravity  In a Plasma Medium}
In this section we study the weak gravitational lensing generated by new massive conformal gravity in the presence of plasma medium.
The refractive index $n(r)$, for new massive conformal gravity is written as
\begin{equation}
n(r)=\sqrt{1-\frac{\omega_{e}^{2}A(r)}{\omega_{\infty}^{2}}},
\end{equation}
their equivalent optical metric, calculated as
\begin{eqnarray}\nonumber
dt^{2}&=&g^{opt}_{ij}dx^{i}dx^{j}=n^{2}(r)[-A(r)dt^{2}\\ \nonumber
&+&dr^{2}B(r)^{-1}+{r}^{2}(d\theta^{2}+\sin^{2}\theta d\phi^{2})].
\end{eqnarray}
In the above expression $A(r)$ and $ B(r)$ can be defined as
\begin{equation}
A(r)=1-\frac{2M}{r}+\frac{Q_{s}^2}{r^2}+\frac{Q_{s}^{2}(-M^2+Q_{s}^2+\frac{6}{m_{2}^2})}{3r^4}+..\nonumber\\
\end{equation}
\begin{equation}
B(r)=1-\frac{2M}{r}+\frac{Q_{s}^2}{r^2}+\frac{2\underline{}Q_{s}^{2}(-M^2+Q_{s}^2+\frac{6}{m_{2}^2})}{3r^4}+..\nonumber\\
\end{equation}
For optical metric the christopher symbols are:
\begin{eqnarray}\nonumber
\Gamma^0_{11}&=&(\frac{B(r)r^{2}}{2})\frac{\omega_{e}^{2}A^{\prime}}{\omega_{\infty}^{2}}(1+\frac{\omega_{e}^{2}A(r)}{\omega_{\infty}^{2}})\\ \nonumber
&+&\frac{A(r)^{\prime}B(r)r^{2}}{2A(r)}-r B(r),
\end{eqnarray}
\begin{eqnarray}\nonumber
\Gamma_{01}^{1}&=&\frac{1}{r}-\frac{A(r)\prime}{2A(r)}
-\frac{\omega_{e}^{2}A^{\prime}}{2\omega_{\infty}^{2}}(1+\frac{\omega_{e}^{2}A(r)}{\omega_{\infty}^{2}}),
\end{eqnarray}
\begin{eqnarray}\nonumber
\Gamma^0_{00}&=&-\frac{\omega_{e}^{2}A^{\prime}}{2\omega_{\infty}^{2}}(1+\frac{\omega_{e}^{2}A(r)}{\omega_{\infty}^{2}})
-\frac{B(r)^\prime}{2B(r)}-\frac{A(r)^\prime}{2A(r)}.
\end{eqnarray}
Now we use the above non-zero christopher symbols and gaussian curvature is calculated as:
\begin{eqnarray} \nonumber
\mathcal{K}&=&\bigg(\frac{2M}{r^{3}}-\frac{6Q_{s}^{2}}{r^{4}}-\frac{29Q_{s}^{2}M^{2}}{3r^{6}}+\frac{7MQ_{s}^{2}}{2r^{5}}\\ \nonumber
&+&\frac{5MQ_{s}^{2}}{r^{7}m^{2}_{2}}+\frac{9M^{2}}{4r^{4}}\bigg)+\frac{\omega_e^2}{\omega_\infty^2}\bigg(\frac{4M}
{r^{3}}+\frac{9Q^{2}}{2r^{4}}\\ \nonumber
&-&\frac{151M^{2}Q_{s}^{2}}{3r^{6}}-\frac{33M^{2}}{r^{4}}-\frac{12MQ_{s}^{2}}{r^{5}}\\ \nonumber
&+&\frac{5MQ_{s}^{2}}{2r^{6}}+\frac{M^{2}}{r^{5}}-\frac{3M^{2}}{2r^{4}}+\frac{6M^{2}Q_{s}^{2}}{r^{7}}\bigg)\\ \nonumber
&+&\frac{\omega_e^4}{\omega_\infty^4}
\bigg(\frac{83M^{2}}{2r^{4}}-\frac{68MQ_{s}^{2}}{r^{5}}+\frac{165M^{2}Q_{s}^{2}}{r^{6}}\\  \nonumber
&+&\frac{3Q_{s}^{2}M^{2}}{r^{8}m^{2}_{2}}-\frac{6M}{r^{3}}+\frac{9Q_{s}^{2}}{r^{4}}+\frac{2M^{2}}{r^{5}}\\
&+&\frac{24M^{2}Q_{s}^{2}}{r^{7}}-\frac{5MQ_{s}^{2}}{r^{6}}\bigg).   \label{AH6}
\end{eqnarray}
Thus, the deflection angle for plasma we apply GBT to acquire it. The deflection angle in weak fields for light beams
since they go straight forward, we apply the constraint that $ r=\frac{b}{sin\varphi}$ at $0^{th}$ order.
\begin{equation}
\alpha=-\lim_{R\rightarrow 0}\int_{0} ^{\pi} \int_\frac{b}{\sin\phi} ^{R} \mathcal{K} dS.
\end{equation}
 After some straightforward calculations the deflection angle for the plasma medium is calculated as
\begin{eqnarray} \nonumber
\alpha&=&\bigg(\frac{3Q_{s}^{2}\pi}{2b^{2}}-\frac{4M}{b}+\frac{29M^{2}Q_{s}^{2}\pi}{32b^{4}}-\frac{14MQ_{s}^{2}}{9b^{3}}\\ \nonumber
&-&\frac{61MQ_{s}^{2}}{60b^{5}m^{2}_{2}}-\frac{5M^{2}\pi}{16b^{2}}\bigg)+\frac{\omega_e^2}{\omega_\infty^2}\bigg(\frac{8M}{b}-\frac{9Q_{s}^{2}\pi}{8b^{2}}\\ \nonumber
&+&\frac{151M^{2}Q_{s}^{2}\pi}{32b^{4}}+\frac{33M^{2}\pi}{4b^{2}}+\frac{16MQ_{s}^{2}}{3b^{3}}\\ \nonumber &-& \frac{15MQ_{s}^{2}\pi}{64b^{4}}
-\frac{4M^{2}}{9b^{3}}+\frac{3M^{2}\pi}{8b^{2}}-\frac{61M^{2}Q_{s}^{2}}{50b^{5}}\bigg)\\ \nonumber &+&\frac{\omega_e^4}{\omega_\infty^4}\bigg(-\frac{83M^{2}\pi}{8b^{2}}
+\frac{272MQ_{s}^{2}}{9b^{3}}+\frac{12M}{b}\\  \nonumber &-&\frac{5M^{2}Q_{s}^{2}\pi}{4b^{6}m^{2}_{2}}-\frac{495M^{2}Q_{s}^{2}\pi}{32b^{4}}-\frac{9Q_{s}^{2}\pi}{4b^{2}}-\frac{8M^{3}}{9b^{3}}\\  \label{A2} &-&\frac{183M^{2}Q_{s}^{2}}{25b^{5}}+\frac{15MQ_{s}^{2}\pi}{32b^{4}}\bigg)
+\mathcal{O}(M^3,Q_{s}^3,b^3).
\end{eqnarray}
The results listed above confirm that light waves are entering the medium of plasma that is homogeneous. The equation  Eq.(\ref{A2}) was simplified to the equation  Eq.(\ref{A1}), as plasma effect can be eliminated. The frequency of photons affects how much the gravitational deflection angle rises due to the presence of plasma medium. In a homogenous plasma medium, photons with a lower frequency or longer wavelength are bent away from the gravitational center at a higher angle. For $\omega\longrightarrow\omega_{e}$, the gravitational deflection angles for longer wavelengths are significantly different which is only feasible for radio waves. The gravitational lens in plasma acts like a radio spectrometer in this sense \cite{1.4}.
\section{Graphical Influence on Deflection angle new Massive conformal Gravity for plasma medium}
This part of the article describe the graphical behavior of deflection of light $\alpha$
regarding $b$ for plasma medium. To avoid the complexity we suppose $Q_{s}=Q$, $\frac{\omega_e^2}{\omega_\infty^2}=\beta$ and $m_{2}=m$.
\begin{figure}
\includegraphics[width=16pc]{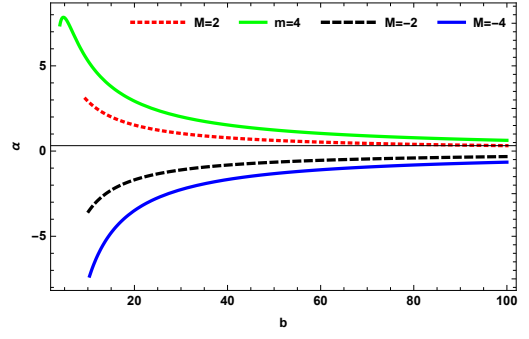}
\caption{\label{label} $\alpha$ versus $b$ with fix value of $m$, $Q$ and $\beta$ for plasma medium.}
\end{figure}
\begin{figure}
\includegraphics[width=16pc]{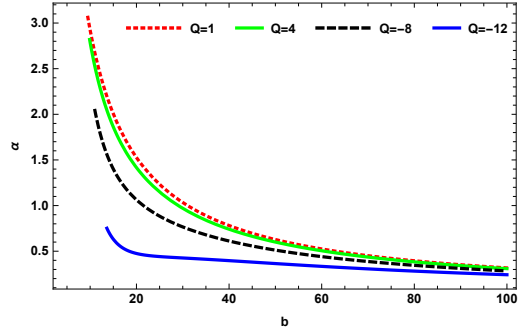}
\caption{\label{label} $\alpha$ versus $b$ with fix value of $m$, $M$ and $\beta$ for plasma medium.}
\end{figure}
In Fig. 3, we show the influence of deflection angle regarding $b$ by taking the fix values of $Q=1$, $m=1$, $\beta=1$ and by varying $M$. For $M>0$ and $M<0$ we examine that the deflection angle indicates the increasing and decreasing behavior respectively.
In Fig. 4, we show the impact of deflection angle versus
$b$ by varying the $Q$ and by taking $M=2$, $m=1$ and $\beta=1$ fixed. We
analyzed that the deflection angle exponentially increased for $Q>0$ and $Q<0$. And deflection angle shows the decreasing behavior for smaller values of $Q$.
\begin{figure}
\includegraphics[width=16pc]{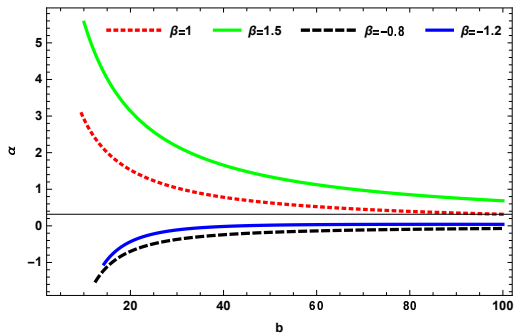}
\caption{\label{label} $\alpha$ versus $b$ with fix value of $m$, $M$ and $Q$ for plasma medium.}
\end{figure}
In Fig. 5, we show the behavior of deflection angle versus $b$ by using the fixed value of $M=2$, $m=1$, $Q=1$ and altering the value of $\beta$. We noticed that for $\beta>0$, deflection angle exponentially increases and on the other hand, for the values of $\beta<0$ the behavior of
deflection angle decreases.
\section{Conclusion}

The present article is about the investigation of the deflection angle in the scenario of non plasma and the plasma medium by the new massive conformal gravity. Firstly, by applying the GBT to obtain the deflection angle of photon and also we analyze the weak gravitational lensing in Eq.(\ref{A1}). We observed that, the  gravitational lensing causes several impacts including amplification, altered image, distortion, and time delay. Present days gravitational lensing is a significant astrophysical device for discover of in accessible objects, a dispersion of dim issue and higher scale structure, checking of the general relativity disclosure of the globe and current astronomical microwave \cite{25}.

Light beams travel through plasma in space. In plasma medium, photons go through in numerous impacts like refraction and dispersing. Change the point of redirection of a light beam which is generally utilized in the estimation of mathematical optics through gravitational lensing. We additionally talk about the graphical influence of various parameters on the deflection angle by new massive conformal gravity. We examined presence of plasma impact on different impacts of gravitational lensing. We described that deflection angle in the influence of the plasma medium defined in  Eq.(\ref{A2}). Here, it must be noted that by neglecting certain boundaries the calculated deflection angle changed over into the BBMB deflection angle \cite{86c}. The plasma effect can vanish if the expression $\frac{\omega_{e}}{\omega_{\infty}}\rightarrow0$ is ignored. We showed how to graphically analyze the deflection angle for new massive conformal gravity in a plasma medium. In addition, we also compared  our graphical results with galactic super-massive black hole \cite{16,18*}. While gravitational lensing with Schwarzschild black hole with respect to solid field breaking point and twisting edge was additionally figured closely resembling the weak field restrictions.\\

\section*{Declaration of competing interest}
The authors declare that they have no known competing financial interests or personal relationships that could have appeared to influence the work reported in this paper.
\section*{Data Availability Statement}
This manuscript has no associated data or the data will not be deposited. There is no observational data related to this article. The necessary calculations and graphic discussion can be made available on request.
\section*{Acknowledgement}
This project was supported by the natural science foundation of China (Grant No. 11975145) and partially supported by JSPS KAKENHI (Grant. No. JP21K03547). The authors thank the reviewers for their comments on this paper.


\vspace{0cm}
\begin{thebibliography}{100}
\bibitem{1} A. Einstein and N. Rosen, Phys. Rev. \textbf{48}, 73 (1935).
\bibitem{2} R. W. Fuller and J. A. Wheeler, Phys. Rev. \textbf{128}, 919 (1962).
\bibitem{3} C. W. Misner and J. A. Wheeler, Ann. Phys.  \textbf{2}, 525 (1957).
\bibitem{4} M. S. Morris, K. S. Thorne and U. Yurtsever, Phys. Rev. Lett. \textbf{61}, 1446 (1988).
\bibitem{4*} S. Nojiri,  S. D. Odintsov and V. Oikonomou, Phys.  Rep. \textbf{692}, 104 (2017).
\bibitem{4**} D. Felice and S. Tsujikawa, Liv.  Rev. in Relat. \textbf{13}, 161 (2010).
\bibitem{5} M. Visser, Phys. Rev. D \textbf{39}, 3182 (1989).
\bibitem{6} M. Visser, American Ins. of Phys. (1995).
\bibitem{1.1} Faria, Advan. in High Ener. Phys. FF3293823 (2014).
\bibitem{1.2} M. Y. Soo, Phys. Lett. B \textbf{730}, 135 (2014).
\bibitem{1.3} Z. D. Cheng and M. Y. Soo, Phys.  Rev. D \textbf{101},  084021 (2020).
\bibitem{1.5}  K. Jusufi and A. \"{O}vg\"{u}n,  Phys. Rev. D \textbf{97}, 024042 (2018).
\bibitem{1.6}  K. Y. Paul, T. Emmanuel and  A. \"{O}vg\"{u}n,  Ann. of Phys. \textbf{436}, 168722 (2022).
\bibitem{1.7} W. Javed, J. Abbas and A. \"{O}vg\"{u}n,  The Eur. Phys. J. C \textbf{79}, 694 (2019).
\bibitem{1.8}  K. Jusufi, C. Marcus and A. \"{O}vg\"{u}n,  Phys. Rev. D \textbf{95}, 104012 (2017).
\bibitem{1.9} K. Yashmitha and A. \"{O}vg\"{u}n,  Chin. Phys. C \textbf{44},  025101 (2020).
\bibitem{1.10} Pantig, C. Reggie and A. \"{O}vg\"{u}n,  Ann. of Phys. \textbf{448}, 169197 (2023).
\bibitem{1.11} A. \"{O}vg\"{u}n, Phys. Rev. D \textbf{98}, 044033 (2018).
\bibitem{1.12} L. Zonghai and A. \"{O}vg\"{u}n,  Phys. Rev. D \textbf{101}, 024040 (2020).
\bibitem{1.13} O. Mert and A. \"{O}vg\"{u}n,  J. of Cosm. and Astro. Phys. \textbf{9}, 2022 (2022).
\bibitem{1.14}  A. \"{O}vg\"{u}n,  Univ. \textbf{5}, 115 (2019).
\bibitem{7} E. Gravanis and S. Willison, Phys. Rev. D \textbf{75}, 084025 (2007).
\bibitem{8} L. Chetouani and G. Cllement, Gen. Relativ. Grav. \textbf{16}, 111 (1984).
\bibitem{9} G. Cllement, Int. J. Theor. Phys. \textbf{23}, 335 (1984).
\bibitem{9*} S. Nojiri, S. D. Odintsov and V. K. Oikonomou. Phys. Rep. \textbf{59}, (2011).
\bibitem{kb1} T. Padmanabhan,  Gen. Relat. and Grav.  \textbf{ 40}, 564 (2008).
\bibitem{kb2} E. J. Copeland, M. Sami and S. Tsujikawa,  Int. J. of Mod. Phys. D \textbf{15}, 1935 (2006).
\bibitem{kb3} T. Clifton, P. G. Ferreira,  A. Padilla and C. Skordis, \textbf{513},  189 (2012).
\bibitem{10} V. Perlick, Phys. Rev. D \textbf{69}, 064017 (2004).
\bibitem{11} M. Safonova, D. F. Torres and G. E. Romero, Phys. Rev. D \textbf{65}, 023001 (2001).
\bibitem{12} A. A. Shatskii, Astron. Rep. \textbf{ 48}, 525 (2004).
\bibitem{13} K. K. Nandi, Y. Z. Zhang and A. V. Zakharov, Phys. Rev. D \textbf{74}, 024020 (2006).
\bibitem{14} A. Bhattacharya and A. A. Potapov, Mod. Phys. Lett. A \textbf{25}, 2399 (2010).
\bibitem{15} F. Abe, Astrophys. J. \textbf{725}, 787 (2010).
\bibitem{15*} A. D. Felice and S. Tsujikawa, Living Rev. Rel. \textbf{3} 13 (2010).
\bibitem{15**}  K. Bamba and S. D. Odintsov,  Symmetry. \textbf{7}, 240 (2015).
\bibitem{16} Y. Toki, T. Kitamura, H. Asada and F. Abe, Astrophys. J. \textbf{740}, 121 (2011).
\bibitem{17} N. Tsukamoto and T. Harada, Phys. Rev. D \textbf{87}, 024024 (2013).
\bibitem{18} T. Kitamura, K. Nakajima  and H. Asada, Phys. Rev. D \textbf{87}, 027501 (2013).
\bibitem{18*} C. Salvatore, Phys. Rept. \textbf{ 509}, 321 (2011).
\bibitem{32} K. Jusufi, I. Sakalli and A. \"{O}vg\"{u}n, Phys. Rev. D \textbf{96}, 024040 (2017).
\bibitem{19} T. Ohgami and N. Sakai, Phys. Rev. D \textbf{91}, 124020 (2015).
\bibitem{20} K. Jusufi, A. \"{O}vg\"{u}n  and A. Banerjee, Phys. Rev. D \textbf{96}, 084036 (2017).
\bibitem{21} H. Asada, Mod. Phys. Lett. A \textbf{32}, 1730031 (2017).
\bibitem{22} T. K. Dey and S. Sen, Mod. Phys. Lett. A \textbf{23}, 953 (2008).
\bibitem{23} K. Nakajima and H. Asada, Phys. Rev. D \textbf{85}, 107501 (2012).
\bibitem{24} K. Jusufi and A. \"{O}vg\"{u}n, Phys. Rev. D \textbf{97}, 024042 (2018).
\bibitem{25} G. W. Gibbons and M. C. Werner, Class. Quant. Grav. \textbf{25}, 235009 (2008).
\bibitem{25*} R. Durrer and R. Maartens, Gen.  Relat. and Grav. \textbf{40}, 328 (2008).
\bibitem{26} M. C.  Werner, Gen. Relat. Grav. \textbf{44}, 3047 (2012).
\bibitem{28} L. Zonghai and J. Jia, The Eur. Phys. J. C \textbf{80}, 157 (2020).
\bibitem{30} K. Jusufi and A. Ovgun, Phys. Rev.  D \textbf{97}, 124024 (2018).
\bibitem{31} K. Jusufi, Gen. Relat. Grav. \textbf{47}, 124 (2015).
\bibitem{33} K. Nozari and S. H. Mehdipour, Europhys. Lett. \textbf{84}, 20008 (2008).
\bibitem{34} C. Gabriel, E. Gallo and A. Rogers, Phys. Rev. D  \textbf{99},  124001 (2019).
\bibitem{35} K. Jusufi and A. \"{O}vg\"{u}n, Phys. Rev. D \textbf{97}, 024042 (2018).
\bibitem{39} K. Nozari and  S. H. Mehdipour,  Eur. phys. Lett. \textbf{84}, (2008).
\bibitem{40} C. Gabriel, E. Gallo and A. Rogers, Phys. Rev. D \textbf{97}, 124016 (2018).
\bibitem{41} E. W. Maxwell,  arXiv:2001.01351v1 [gr-qc].
\bibitem{42} S. Wienberg, New York, (1972).
\bibitem{43} G. S. Bisnovatyi-Kogan and O. Y. Tsupko, Grav. Cosmol. \textbf{20}, 220 (2014).
\bibitem{44} O. Y. Tsupko and G. S. Bisnovatyi-Kogan, Phys. Rev. D \textbf{87}, 124009 (2013).
\bibitem{45} G. Crisnejo and E. Gallo, Phys. Rev. D \textbf{97}, 124016 (2018).
\bibitem{46} W. Javed, J. Abbas and A.\"{O}vg\"{u}n, Phys. Rev. D \textbf{100}, 044052 (2019).
\bibitem{46a} M. Y. Soo and D.C. Zou. Phys. Rev. D \textbf{ 100}, 064057 (2019).
\bibitem{47} A. Rogers, Mon. Not. Roy. Astron. Soc. \textbf{451}, 17 (2015).
\bibitem{48} V. Morozvan,  B. Ahmadov and A. Tursunov, Astro.  space sci. \textbf{346}, 513 (2013).
\bibitem{50}  M. Sharif and W. Javed,  Eur. Phys. J. C  \textbf{72}, (2012).
\bibitem{51} H. S. Vieira, V. B. Bezerra and C. R. Muniz, Ann.  Phys. \textbf{350}, (2014).
\bibitem{53} G. W. Gibbons and C. M. Warnick, Phys. Rev. D \textbf{79}, 064031 (2009).
\bibitem{54} G. W. Gibbons et al, Phys. Rev. D \textbf{79}, 044022 (2009).
\bibitem{55} G. W. Gibbons and M. Vyska, Class. Quant. Grav. \textbf{29}, 065016 (2012).
\bibitem{56} C. Bloomer, arXiv:1111.4998 [math-ph].
\bibitem{57} G. W. Gibbons, Class. Quant. Grav.  \textbf{33}, 025004 (2016).
\bibitem{58} R. S. Pdas and S. Ghosh, Eur. Phys. J. C \textbf{77}, 735 (2017).
\bibitem{59} I. Sakalli and A. \"{O}vg\"{u}n, EPL. \textbf{118}, 60006 (2017).
\bibitem{60} K. Jusufi, M. C. Werner, A. Banerjee and A. \"{O}vg\"{u}n, Phys. Rev. D \textbf{95}, 104012 (2017).
\bibitem{61} T. Ono, A. Ishihara and H. Asada, Phys. Rev. D \textbf{96}, 104037 (2017).
\bibitem{62} K. Jusufi, I. Sakalli and A. \"{O}vg\"{u}n, Phys. Rev. D \textbf{96}, 024040 (2017).
\bibitem{63} H. Arakida, Gen. Rel. Grav. \textbf{50}, 48 (2018).
\bibitem{64} A. \"{O}vg\"{u}n, I. Sakalli and J. Saavedra, JCAP. \textbf{1810}, 041 (2018)
\bibitem{65} A. \"{O}vg\"{u}n, I. Sakalli and J. Saavedra, Ann. Phys. \textbf{411}, 167978 (2019).
\bibitem{66} A. \"{O}vg\"{u}n, G. Gyulchev and K. Jusufi, Ann. Phys. \textbf{406}, 152 (2019).
\bibitem{67} A. \"{O}vg\"{u}n, K. Jusufi and I. Sakalli, Ann. Phys. \textbf{399}, 193 (2018).
\bibitem{69} A. \"{O}vg\"{u}n, Phys. Rev. D \textbf{98}, 044033 (2018).
\bibitem{70} T. Ono, A. Ishihara and H. Asada, Phys. Rev. D \textbf{99},  124030 (2019).
\bibitem{71} G. Crisnejo, E. Gallo and A. Rogers, Phys. Rev. D \textbf{99}, 124001 (2019).
\bibitem{72} Z. Li and  A. \"{O}vg\"{u}n, doi:\textbf{10.20944}/preprints201911.0195.v1.
\bibitem{73} Z. Li and T. Zhou, arXiv:1908.05592 [gr-qc].
\bibitem{74} Z. Li and J. Jia, arXiv:1912.\textbf{05194} [gr-qc].
\bibitem{75} Z. Li, G. He and T. Zhou, arXiv:1908.\textbf{01647} [gr-qc].
\bibitem{76} A. \"{O}vg\"{u}n, Universe. \textbf{5}, 115 (2019).
\bibitem{77} W. Javed, R. Babar and A. \"{O}vg\"{u}n, Phys. Rev. D \textbf{99}, 084012 (2019).
\bibitem{79} A. \"{O}vg\"{u}n, Phys. Rev. D \textbf{99}, 104075 (2019).
\bibitem{80} K. Jusufi and A. \"{O}vg\"{u}n, Int. J. Geom. Meth. Mod. Phys. \textbf{16},  1950116 (2019).
\bibitem{81} W. Javed, R. Babar and A. \"{O}vg\"{u}n, Phys. Rev. D \textbf{100}, 104032 (2019).
\bibitem{82} Y. Kumaran and A. \"{O}vg\"{u}n, arXiv:1905.\textbf{11710} [gr-qc].
\bibitem{82*} C. Y. Fu and C. Salvatore,  Prog. Phys. \textbf{79}, 106901 (2016).
\bibitem{82**} K. Bamba et al,  Astro. phys. and Space Sci. \textbf{342}, 228 (2012).
\bibitem{84} A. \"{O}vg\"{u}n, I. Sakalli and J. Saavedra, arXiv:1908.04261 [gr-qc].
\bibitem{85} K. Jusufi, A. \"{O}vg\"{u}n, A. Banerjee  and I. Sakalli, Eur. Phys. J. Plus. \textbf{134}, 428 (2019).
\bibitem{kb86} G. Crisnejo and  E. Gallo, Phys.  Rev. D \textbf{97}, 124016 (2018).
\bibitem{1.4} B. Kogan and O. Y.  Tsupko, Grav. and Cosm. \textbf{15}, 27 (2009).
\bibitem{86c} W. Javed and A. \"{O}vg\"{u}n, Eur. Phys. J. Plus. \textbf{135}, 6 (2020).
\end{thebibliography}
\end{document}